\documentclass{Interspeech2024}
\usepackage[nameinlink,capitalize]{cleveref}
\usepackage{multirow}
\usepackage{cite}
\usepackage{braket}

\interspeechcameraready

\title{SOT Triggered Neural Clustering for Speaker Attributed ASR}

\name[affiliation={1*}]{Xianrui}{Zheng}
\name[affiliation={1}]{Guangzhi}{Sun}
\name[affiliation={2}]{Chao}{Zhang}
\name[affiliation={1}]{Philip C.}{Woodland}

\address{
  $^1$Cambridge University Engineering Dept., Trumpington St., Cambridge, CB2 1PZ U.K\\$^2$Department of Electronic Engineering, Tsinghua University
}
\email{$^1$\{xz396, gs534, pw117\}@cam.ac.uk, $^2$cz277@tsinghua.edu.cn}


\begin{document}

\maketitle

\blfootnote{$^*$Supported by an Amazon Studentship. \\
This work has been performed using resources provided by the Cambridge Tier-2 system operated by the University of Cambridge Research Computing Service (www.hpc.cam.ac.uk) funded by EPSRC Tier-2 capital grant EP/T022159/1}

\begin{abstract}
This paper introduces a novel approach to speaker-attributed ASR transcription using a neural clustering method. 
With a parallel processing mechanism, diarisation and ASR can be applied simultaneously, helping to prevent the accumulation of errors from one sub-system to the next in a cascaded system. 
This is achieved by the use of ASR, trained using a serialised output training method, together with segment-level discriminative neural clustering (SDNC) to assign speaker labels. 
With SDNC, our system does not require an extra non-neural clustering method to assign speaker labels, thus allowing the entire system to be based on neural networks. 
Experimental results on the AMI meeting dataset demonstrate that SDNC outperforms spectral clustering (SC) by a 19\% relative diarisation error rate (DER) reduction on the AMI Eval set. 
When compared with the cascaded system with SC, the parallel system with SDNC gives a 7\%/4\% relative improvement in cpWER on the Dev/Eval set.

\end{abstract}

\section{Introduction}

The task of identifying `who spoke what' is commonly achieved through cascaded systems \cite{raj2021integration,watanabe2020chime,zheng2022tandem}. Initially, a voice activity detection (VAD) system is applied to filter out non-speech regions from the entire recording, with the speech regions being the VAD segments between non-speech regions. 
A speaker diarisation system then determines `who spoke when' by breaking the VAD segments into speaker homogeneous segments. 
The segmented output is then processed by a speaker recognition system that determines the content spoken by each speaker. 
In situations where speech from different speakers overlap, a source separation system  \cite{raj2023gpu} can be employed before the recognition system to disentangle overlapping speech into separate streams.
This multi-stage cascaded framework, however, is less than ideal. 
Errors produced in early stages can propagate through to subsequent stages, often compounding inaccuracies. 
Correction of these errors by later stages is challenging because the input of the later stage is purely based on the output of the previous stage. 
Current research is trying to develop more integrated architectures that can reduce error accumulation. 

To avoid the transmission of errors from the diarisation stage to the automatic speech recognition (ASR) stage, a framework known as serialised output training (SOT) \cite{kanda2020serialized} has been introduced.
SOT processes entire VAD segments, rather than taking the speaker homogeneous segments from the diarisation system. 
SOT has the capability to detect the presence of multiple speakers within each VAD segment. It then sequentially generates transcriptions, starting with the first speaker in the segment, followed by the second speaker, and continues this pattern for any additional speakers.

Several extensions to the SOT ASR have been explored to also provide speaker information \cite{kanda2021end,kanda2022transcribe,cornell2024one}.
However, they still rely on a non-neural clustering method after SOT to assign speaker labels across an entire meeting. 
In this paper, we propose a neural clustering method, called segment-level discriminative neural clustering (SDNC), to work together with SOT to directly assign relative speaker labels across the entire meeting, removing the need to use an extra non-neural clustering method. 
In our framework, SOT identifies the number of speakers within each VAD segment, while the SDNC generates matching speaker labels. 
This method avoids errors that typically accumulate when data passes from diarisation to ASR.
Experimental results on the augmented multi-party (AMI) meeting dataset show that the proposed SDNC is able to outperform spectral clustering in terms of DER and cpWER. 

The paper is organised as follows. 
Section 2 discusses related work. 
Section 3 presents SDNC and describes how it is used together with SOT. 
The experimental setup is described in Sec. 4 and Sec. 5 gives results, followed by conclusions.

\section{Related Work}

Unsupervised non-neural clustering methods have been widely used for the speaker diarisation task \cite{romero2017speaker,dawalatabad2021ecapa,sun2021combination,landini2022bayesian,xia2022turn}, with speaker embeddings trained via neural networks. 
Recently, several methods have been proposed to perform supervised training for the speaker diarisation task \cite{zhang2019fully,medennikov2020target,li2021discriminative,wang2023target,chen2023attention,10094752,Yousefi2023speaker,horiguchi2022encoder,Fujita2019Interspeech,kinoshita2021eendvc}.
EEND-based methods \cite{Fujita2019Interspeech,horiguchi2022encoder,kinoshita2021eendvc,khare2022asr} were proposed to perform frame level speaker label assignment, without any prior knowledge of target speakers \cite{medennikov2020target,Yousefi2023speaker}. 
Although EEND methods assign relative speaker labels to a stream of audio, they cannot handle long meetings due to computational memory constraints. Therefore, there is still a necessity to collect speaker embeddings and employ additional non-neural clustering methods to assign speaker labels for long meetings.
In order to process the entire meeting in a single step, rather than using the entire audio to provide frame-level clustering results, speaker embeddings extracted from short audio segments can be used as inputs \cite{zhang2019fully,li2021discriminative}. 
The output will then be the clustering results for each speaker embedding.

Discriminative neural clustering (DNC) \cite{li2021discriminative} employs a transformer-based encoder-decoder structure, with a constraint such that the length of the input sequence to the encoder is the same as the length of the output sequence from the decoder. 
The input to the encoder is a sequence of speaker embeddings, and the decoder outputs a sequence of relative speaker labels. 
Let the input vector sequence be $\bm{X} = [\bm{x}_1, \dots, \bm{x}_N]$,
and the output cluster label sequence be $\bm{C} = [\bm{c}_1, \dots, \bm{c}_N]$. 
The first cluster $\bm{c}_1$ is always given an ID of 1, and if the model identifies another cluster in the sequence, that cluster will have an ID of 2 \textit{etc}. 
The speaker embeddings were first extracted using a fixed-length window sliding through each utterance, utilising oracle utterance boundary information. 
To assign a single speaker label per utterance, the speaker embeddings corresponding to an utterance are first averaged before being fed into DNC. 

Two systems \cite{kanda2022transcribe,cornell2024one} that avoid the use of a cascaded framework, without target speaker profiles, and employ SOT are worth mentioning. 
\cite{kanda2022transcribe} transcribes words in a serialised format, including tokens to indicate speaker change. 
The original SOT model, which utilises a conventional Attention Encoder Decoder (AED) framework, is designed to transcribe words from different speakers within the current VAD segment in a serialised manner \cite{kanda2020serialized}.
However, \cite{kanda2022transcribe} introduces an additional encoder and decoder into the process. This extra component generates a speaker embedding for each ASR output token. 
Another work \cite{cornell2024one} combines SOT with an EEND based model\cite{kinoshita2021eendvc}.
Unlike methods that only predict the speaker change token, the output of the ASR decoder is also responsible for predicting a start and end time token and a local speaker ID for each utterance in the input segment. 
The predicted local speaker IDs are linked with their corresponding local speaker vectors. 
These vectors for each local speaker ID are then averaged to perform clustering later for the assignment of global speaker IDs. 

These two recent works give, to the best of our knowledge, the best cpWER \cite{watanabe2020chime} results on AMI. 
However, they both require training the model to provide speaker embeddings and ASR outputs simultaneously. 
To perform such a joint task, extensive supervised training data for ASR is required. 
Besides AMI, \cite{kanda2022transcribe} uses 75k hours of in-house data to train ASR, and then transcribe VoxCeleb 1\&2 \cite{nagrani2020voxceleb,chung2018voxceleb2} data for training their joint system. 
\cite{cornell2024one} uses 5k hours of simulated meetings from SINS \cite{dekkers2017sins} and LibriSpeech \cite{panayotov2015librispeech}, and around 100 hours of data from Mixer 6 \cite{brandschain2010mixer} and CHiME-6 \cite{watanabe2020chime}. 
Additionally, the assignment of meeting-level speaker IDs requires the use of a non-neural algorithm for transcribing long meetings. 

In this paper, we aim to use a neural clustering algorithm to directly assign meeting-level speaker labels without target speaker profiles. 
To simplify the whole training procedure, the only supervised data we use to train our ASR and neural clustering is AMI.

\section{Methodology}

Segment-level discriminative neural clustering (SDNC) is an extension of the original DNC \cite{li2021discriminative}.
The distinction between DNC and SDNC is their input-output relation: for DNC, each speaker embedding in the input is assigned a cluster label, and the number of inputs and outputs is equal. This one-to-one correspondence does not apply to SDNC.
DNC does not aim to identify multiple speakers in overlap regions, but SDNC is able to give multiple speaker labels for overlap regions. 
Unlike DNC, where the window-level speaker embeddings are extracted for each utterance and then averaged before being fed into the encoder,
SDNC directly takes the windows sliding across the VAD segment, \textit{i.e.} speech regions between two non-speech regions.

\subsection{Model details for SDNC}

\begin{figure}[ht]
    \centering
    \includegraphics[width=.9\linewidth]{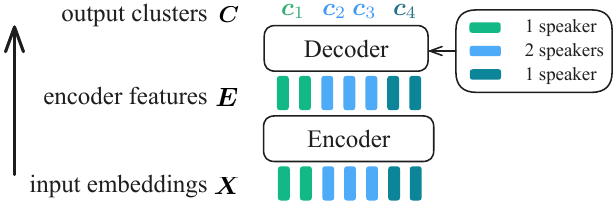}
    \vspace{-2mm}
    \caption{A toy example of the input and output of the SDNC. Different colours represent different VAD segments and multiple speaker clusters can be output in one VAD segment with SDNC.}
    \label{fig:SDNC}
\end{figure}

The input to the encoder is a sequence of window level speaker embeddings $\bm{X} = [\bm{x}_1, \dots, \bm{x}_N]$, each of them are from one of the $M$ VAD segments. 
The output of the encoder will be $\bm{E} = [\bm{e}_1, \dots, \bm{e}_N]$.
The decoder takes the encoder outputs, together with external information regarding the number of speakers in each segment to decide the length of the output sequence.
For each output label, the transformer decoder cross attention will attend to the encoder output features corresponding to the current segment only, while
all encoder features corresponding to other segments will be masked out, \textit{i.e.} the cross attention for each output class label belonging to the $i^{\text{th}}$ VAD segment $\bm{V}_i$ only attends to 
\begin{equation}
    \set{\bm{e}_j | \bm{x}_j \in \bm{V}_i}.
\end{equation}
\cref{fig:SDNC} shows an example, where the eight input speaker embeddings are from three VAD segments. 
The external information tells the SDNC decoder that the first segment (in blue) has one speaker the second segment (in yellow) has two unique speakers and the third segment (in grey) has one speaker. 
The SDNC decoder will then predict four outputs in total.
When outputting $\bm{c}_1$, the cross attention of the SDNC decoder will attend to the first two encoder features, 
and attend to the next three encoder features when outputting $\bm{c}_2$ and $\bm{c}_3$. 

\subsection{`First Speaker' segmentation and windows}
\begin{figure}[ht]
\vspace{3mm}
    \centering
    \includegraphics[width=.95\linewidth]{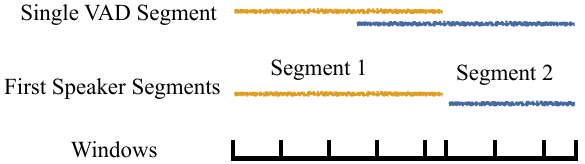}
    \caption{Windows for `First Speaker' segmentation. Different colours represent different speaker turns.}
\vspace{-4mm}
    \label{fig:seg_1}
\end{figure}

Before training SDNC to decode multiple outputs for a VAD segment containing multiple speakers,
SDNC is first pre-trained to decode one output for each segment. 
A VAD segment with overlap will be split into multiple speaker segments. 
As shown in \cref{fig:seg_1}, there are two speakers in the VAD segment with overlapping speech.
The overlapped region is first assigned to the speaker speaking first, creating two `First Speaker' segments, a fixed-size window then slides across each segment. 
The length of the last window for each `First Speaker' segment cannot exceed the end of that segment. 
The target for SDNC is to predict one relative speaker label for each `First Speaker' segment. 
During SDNC fine-tuning, the speaker embeddings are from windows sliding across the entire VAD segment.

\begin{figure*}[ht!]
    \centering
    \includegraphics[width=.99\linewidth]{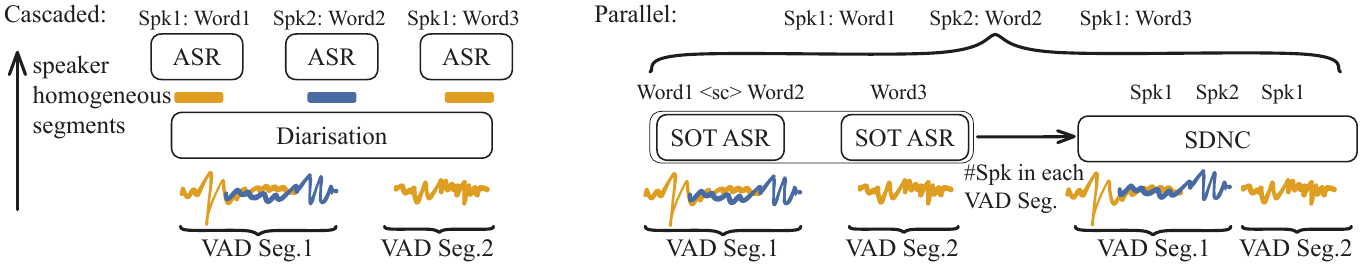}
    \vspace{-1mm}
    \caption{Comparison between the cascaded system (left) and the proposed parallel system (right). Different colours represent different speakers. `$<$sc$>$' represents speaker change. }
    \label{fig:cascadedvsparallel}
    \vspace{-5mm}
\end{figure*}

\subsection{Data augmentation}

Since the input to the model is the entire meeting's speaker embeddings, for real meeting corpora there aren't many data points available for training. 
DNC \cite{li2021discriminative} mainly applied two augmentation methods: 
one is called input vector randomisation and the other is Diaconis augmentation.
We have modified these augmentation techniques in order to achieve better performance. 

The original input vector randomisation method involves sampling speaker embeddings at random to form new input sequences. 
However, this method may select the same speaker embedding multiple times, resulting in less variation within the input vector. 
To address this, we propose shuffling the segments belonging to the same speaker. 
Furthermore, Diaconis augmentation is designed to apply a random rotation matrix to the original speaker embeddings. 
We have implemented this augmentation on-the-fly for all training examples, rather than pre-computing the augmented embeddings, which enables the generation of a greater number of new training examples and reduces memory storage requirements.

\subsection{Parallel System with SOT}
\cref{fig:cascadedvsparallel} shows the differences between the cascaded system and the proposed parallel system. 
The input of the cascaded system for the ASR is the output from the diarisation system.
In the parallel system, each VAD segment is fed into the SOT ASR to decode transcriptions along with speaker change tokens.  
Simultaneously, all speaker embeddings from each VAD segment are fed into the encoder of SDNC. 
SOT then informs the SDNC of the number of speakers present in each VAD segment, enabling the SDNC to decode the appropriate number of speaker labels. Finally, the outputs from both the SOT ASR and SDNC are combined to give the final speaker-attributed ASR output.

\section{Experiment Setup}

\subsection{Dataset: AMI corpus}
AMI is a meeting corpus consisting of 100 hours of recorded meetings.
It has a total of 169 meetings: 135 designated for training, 18 for Dev set, and 16 meetings from the Eval set.
The number of speakers in a meeting is between 3 and 5. 
The multi-distant microphone (MDM) audio is used in our experiments, where the eight-channel arrays were combined using BeamformIt \cite{angueraAcoustic2007}. 
As described in \cite{sun2021combination}, the original manual segmentation marks a lot of non-speech ``silence" regions as speech. 
For diarisation, it is important to properly identify and remove large regions of ``silences''. 
We follow the same procedure as in \cite{sun2021combination} to strip those ``silences'' by using a forced alignment with the reference transcriptions and a pre-existing speech recognition system \cite{youngHTK2015}. 
Compared to the original speech regions, the silence-stripped data reduces the total duration by 9.9\% for Dev and by 11.7\% for Eval.

\subsection{Speaker embeddings, ASR and baseline}
The ECAPA-TDNN \cite{dawalatabad2021ecapa} implementation in \cite{speechbrain} was used to extract speaker embeddings for each window. 
In our experiments, each window has a 1.5 sec. window length and a 1.5 sec. stride. 
Since embeddings from ECAPA were used to train SDNC, the spectral clustering from the ECAPA pipeline in \cite{speechbrain} was chosen to be the baseline for our experiments. 
Other methods such as the ones in \cite{landini2022bayesian,cornell2024one,kanda2022transcribe} are not comparable to our system since they are trained on far more supervised data.
ECAPA embeddings are trained entirely on VoxCeleb 2 \cite{chung2018voxceleb2} data. 
ASR uses the pre-trained wavLM-base-plus encoder \cite{chen2022wavlm} and a 6-layer transformer decoder trained from scratch. 
The AMI MDM corpus is the only supervised data used to train SDNC and ASR.

\subsection{Evaluation metrics}
Evaluation is based on the diarisation error rate (DER), word error rate (WER) and concatenated minimum-permutation
word error rate (cpWER) \cite{watanabe2020chime}.
DER is a time-based metric while WER and cpWER are both word-based. 
WER is calculated for each segment without considering speaker identity, whereas cpWER concatenates transcriptions from the same speaker and finds the minimum WER from all possible speaker mappings between the predicted speaker labels and the reference speaker IDs.
DER uses a 0.25 sec. collar and includes overlap regions. 
cpWER is scored using MeetEval \cite{MIMO23}. 
Sometimes the predicted sequence of speaker labels for VAD segments with overlapping speech may be inaccurate.
For cpWER-P, the predicted speaker labels are permuted to get the minimum cpWER. 
Around 70\% of VAD segments are speaker homogeneous. 
We further score only these segments to obtain DER-H, WER-H and cpWER-H. 

\section{Results}

\subsection{ASR results}

With utterance segmentation as training data in \cref{tab:asr}, the input to ASR is single utterances.
With utterance segmentation as test data, overlapping regions will be processed multiple times since they belong to part of the utterances of different speakers. 
With VAD segmentation as the test data for ASR trained on utterance segmentation, words of overlapping regions are arranged in time order. 
The WER-H in the first line in \cref{tab:asr} is better than the second line because VAD segmentation combines continuous utterances, thus providing more context. 

When the training data utilises VAD segmentation (last line in \cref{tab:asr}), SOT-style training is employed. In this case, for each VAD segment, the transcription of the first speaker is placed before those of subsequent speakers.
When comparing SOT-style ASR to the ASR trained with utterance segmentation, there is an absolute WER reduction of 2\% on Dev and 3\% on Eval, respectively, when VAD segmentation is used for testing.

\begin{table}[]
    \centering
    \caption{WER/WER-H using oracle diarisation with utterance (Utt.) and VAD segmentations.}
    \begin{tabular}{cccccc}
    \toprule
    \multirow{2}{*}{Train} & \multirow{2}{*}{Test} & \multicolumn{2}{c}{WER} & \multicolumn{2}{c}{WER-H} \\
    & &  Dev & Eval & Dev & Eval \\
       \midrule
       Utt. Seg. & Utt. Seg. & 22.4 & 24.7 & 15.9 & 17.1 \\
       Utt. Seg. & VAD Seg. & 27.7 & 29.6 & 15.7 & 16.9 \\
       VAD Seg. & VAD Seg. & 25.8 & 26.6 & 15.4 & 16.9 \\
       \bottomrule
    \end{tabular}
    \label{tab:asr}
\end{table}

\subsection{`First Speaker' Segmentation}

\cref{tab:assign_first_result} shows results with `First Speaker' segmentation.
With this segmentation, the diarisation system only needs to give one speaker label to each `First Speaker' segment, therefore all systems are non-overlap aware.  
The input to the ASR will be the entire `First Speaker' segment. 
The DER for spectral clustering (SC) utilises the ECAPA pipeline as detailed by  \cite{dawalatabad2021ecapa}.
Majority voting is applied to ensure that each `First Speaker' segment is associated with a single estimated speaker.
Now that each `First Speaker' segment possesses both a speaker label and a transcription provided by the ASR, these elements can be combined when calculating cpWER.

SDNC yields a relative DER reduction of 3\% on Dev and 19\% on Eval when compared to SC. 
A Wilcoxon signed-rank test for both Dev and Eval sets gives a p-value of 0.05\% showing that the improvement is statistically significant.
Moreover, there is a relative cpWER reduction of 3\% for Dev and 8\% for Eval. 
Since the decoded words for each `First Speaker' segment are identical for SC and SDNC, the reduction in cpWER is solely attributable to the improved speaker assignment by SDNC.

\begin{table}[]
    \centering
        \caption{DER (DER-H) and cpWER results for non-overlap aware systems. 
        DER in parentheses is DER-H.}
    \begin{tabular}{ccccc}
    \toprule
    \multirow{2}{*}{Method} & \multicolumn{2}{c}{DER (DER-H)} & \multicolumn{2}{c}{cpWER} \\
       & Dev & Eval & Dev & Eval \\
       \midrule
       SC & 5.8 (1.5) & 5.4 (1.8) & 32.2 & 35.1 \\
       SDNC & 5.6 (1.5) & 4.4 (1.1) & 31.1 & 32.4 \\
       \bottomrule
    \end{tabular}
    \label{tab:assign_first_result}
\end{table}

\subsection{VAD Segmentation}

\begin{table}[t]
    \centering
    \caption{DER (DER-H) and cpWER results for overlap-aware systems. SDNC only has DER-H since it does not have times for each cluster label for VAD segments that include overlap.}
    \begin{tabular}{cccccc}
    \toprule
    \multirow{2}{*}{System} &
    \multirow{2}{*}{Method}
    & \multicolumn{2}{c}{DER (DER-H)} & \multicolumn{2}{c}{cpWER} \\
       & & Dev & Eval & Dev & Eval \\
       \midrule
       Cascaded & SC & 8.6 (3.8) & 5.6 (2.1) & 37.6 & 35.9 \\
       Parallel & SDNC & (1.9) & (1.6) & 34.8 & 34.6 \\
       \bottomrule
    \end{tabular}
    \label{tab:cpWERVAD}
\end{table}

\begin{table}[t]
    \centering
    \caption{cpWER-H and cpWER-P results on parallel systems with different clustering methods.}
    \begin{tabular}{ccccc}
    \toprule
\multirow{2}{*}{Method} & \multicolumn{2}{c}{cpWER-H} & \multicolumn{2}{c}{cpWER-P}\\
& Dev & Eval & Dev & Eval \\
 \midrule
SC & 23.6 & 21.2 & 30.0 & 30.3 \\
SDNC & 20.8 & 20.5 & 28.5 & 28.9 \\
 \bottomrule
\end{tabular}
\label{tab:cpWERHP}
\end{table}

The SDNC model used in \cref{tab:cpWERVAD,tab:cpWERHP} was first pre-trained on windows from the `First Speaker' segmentation, and then fine-tuned on windows directly from the oracle VAD segmentation. 
Unlike `First Speaker' segmentation, which assumes that each segment contains only one speaker, VAD segmentation can have multiple speakers within a single segment. 
Therefore, systems may need to assign multiple speaker labels for a segment, making this an overlap-aware setting.
Comparison of \cref{tab:cpWERVAD} and \cref{tab:assign_first_result} with SC shows that on the Dev data, there is a 48\% relative DER and 17\% relative cpWER degradation. 
When using a parallel system with SDNC, the relative cpWER degradation on Dev is 12\%. 
This shows that the removal of the `First Speaker' segmentation affects the performance significantly.

To determine the number of speakers in each VAD segment, 
the number of speaker change tokens from SOT ASR are counted. 
For the parallel system, this number will be used to determine the length of the SDNC output sequence. 
When utilising `First Speaker' segmentation, SDNC assigns one label to each segment, therefore it can calculate the DER by using the start and end times of each segment. 
However, when employing VAD segmentation, SDNC may need to decode multiple labels for a single segment. 
Since it no longer has the start and end times associated with each output, computing DERs for overlapping segments becomes challenging. 
However, DER-H can be still be calculated by using the first decoded label for each segment. 
The comparison in \cref{tab:cpWERVAD} is between cascaded and parallel systems. 
Since SC assigns cluster labels to each window, the cascaded system uses SC to split VAD segments into predicted speaker homogeneous segments. 
Then the resulting segments from SC are decoded by ASR. 
The DER-H for the cascaded system has degraded from below 2\% to over 2\%. In contrast, the DER-H for the parallel system remains below 2\%, and the cpWER exhibits a 7\% relative reduction on Dev and a 4\% reduction on Eval when compared to the cascaded system.

To further compare SC and SDNC, SC was also used in the parallel system. 
For VAD segments that only have a single speaker, majority voting can be applied to ensure only one speaker is assigned to each segment.
However, assigning speaker labels to VAD segments with overlapping speech is not straightforward.
This complexity arises because the number of speakers estimated for such segments may differ between SC and SOT. 
Therefore, when SOT predicts multiple speakers for a VAD segment, the unique speaker labels predicted by SC/SDNC for that segment are permuted to find the sequence that minimises cpWER (cpWER-P). 
\cref{tab:cpWERHP} shows that SDNC performs better than SC for both cpWER-H and cpWER-P, giving a 3\% and 5\% relative Eval cpWER-H and cpWER-P reductions. 
This result shows that the SDNC can predict more accurately than the SC within a parallel system. 

\section{Conclusions}

This paper proposed a neural clustering method that can be used to perform diarisation for long meetings, without using an extra non-neural clustering algorithm.
In addition, this method can be used in parallel with an SOT ASR.
Compared to a cascaded system, the SOT ASR in the parallel system does not need to use the output from the diarisation system, therefore mitigating the accumulation error issue in the cascaded system. 
Experiments conducted on the AMI dataset showed that the proposed parallel system with SDNC performs better than spectral clustering with either a cascaded or parallel system, giving an 8\% and 4\% relative cpWER reduction on the AMI evaluation set under the non-overlap aware and overlap aware settings respectively.

\bibliographystyle{IEEEtran}
\bibliography{mybib}
\end{document}